# The effect of manganese and silicon additions on the corrosion resistance of a polycrystalline nickel-based superalloy

E. Anzini[1], N. Glaenzer[2], P. M. Mignanelli[3], M. C. Hardy[3], H. J. Stone[2], S. Pedrazzini[1]

[1] Department of Materials, Imperial College London, Exhibition Road, SW7 2AZ, London, UK.
[2] Department of Materials Science and Metallurgy, University of Cambridge, 27 Charles Babbage road, Cambridge, CB3 0FS, UK.
[3] Rolls-Royce plc, PO Box 31, Derby DE24 8BJ, UK.

## Abstract
The service lives of nickel superalloys are often limited by environmental degradation. The present study compares oxidation, sulfidation and hot corrosion at 750°C of three variants of a polycrystalline superalloy: a baseline alloy, a variant containing 1wt% Mn and one containing 0.5wt% Si. Mn reduced the oxidation rate without changing the scale morphology. The $MnCr_2O_4$ scale formed proved more protective against sulfidation and hot corrosion, but internal sulfides extended the damage depth. Si modified the oxide morphology to a continuous $Cr_2O_3$-$Al_2O_3$ dual layer. This provided improved protection, reducing the sulfidation depth by 2/3 and the hot corrosion depth by ½.

***Keywords***: Nickel superalloy, Manganese, Silicon, Oxidation, Sulfur, Salt, Corrosion.

## Introduction

Nickel-based superalloys are common structural materials employed in the hottest part of jet engines and industrial gas turbines for power generation. In these extremely harsh operating environments, sulfur contamination can arise from exposure of components to fuel, atmospheric gases and/or exhaust gases in-service. This contamination may cause substantial reductions in the lifetime of nickel-based superalloy components, particularly under creep and dwell fatigue conditions [1]–[4]. The environmental degradation mechanisms involved are generally divided into two main sub-groups: Type I and Type II hot corrosion [5]. Type I hot corrosion occurs above 800-900 °C where sodium sulfate ($Na_2SO_4$) is above its melting temperature. This type of corrosion is known to cause the failure of the protective oxide scale, and in some polycrystalline alloys the formation of Cr-Ti sulfides in the substrate [3]. In contrast Type II hot corrosion occurs at lower temperatures through reactions with substrate elements (eg. $NiSO_4$-$Na_2SO_4$ or $CoSO_4$-$Na_2SO_4$) to form liquid salt phases with low melting points between 560-700°C [5], [6]. One of the key industrial applications of nickel-based superalloys are the turbine disc rotors that form the core of gas turbine engines. These components are typically uncoated and experience temperatures below 800°C in service. As their failure in service cannot be tolerated, it is critical that their resistance to type II hot corrosion is fully understood for their safe application [7].

Mn is a relatively uncommon alloying element in nickel-based superalloys. However, a recent study has highlighted the beneficial effect of Mn on the oxidation resistance of the polycrystalline superalloy V207J, reporting the formation of a thermally grown oxide scale composed of mixed $MnCr_2O_4$ spinel and $Cr_2O_3$ [8]. Unlike Cr, Mn has similar sulfidation and oxidation rates and activation energies [9], [10], but its effect on the hot corrosion resistance of nickel-based superalloys remained unexplored. Manganese sulfides (MnS, $MnS_2$) are stable, slow growing and can have similar growth rates to $Cr_2O_3$. In steels, some of the benefits of MnS formation are usually lost due to the high solubility of Fe in MnS, which allows the passage of Fe though the MnS, leading to the growth of an external FeS layer [11], [12]. Furthermore, Mn in steels has the tendency to form undesirable sulfide inclusions during production processing, which can cause crack initiation. In contrast, the fabrication of aerospace-grade Ni superalloys is typically performed under carefully controlled low-sulfur conditions to reduce defect formation and improve performance. This reduces the need to include Mn in the alloy for the purpose of S scavenging during manufacture.

Unlike Mn, Si is a comparatively common low-level alloying addition in polycrystalline nickel-based superalloys and it has been reported to have a beneficial effect on the oxidation resistance [13]. Specifically, it was shown that Si may promote the formation of a continuous layer of alumina underneath the chromia scale in the polycrystalline superalloy SCA425+ between 900-1000°C [13]. An alumina scale offers advantages as it is generally regarded as being more protective than chromia against sulfur-based corrosion, though the mechanisms require further understanding. In addition, Si has been reported to segregate to grain boundaries within the alumina and aluminium-containing spinels [14]. And this segregation may affect the inward diffusion of sulfur at grain boundaries, further improving sulfidation resistance.

To obtain an improved understanding of the effect of minor additions of Mn and Si, in the present work, the effects of low-level additions of these elements on the sulfidation and hot corrosion resistance of the polycrystalline nickel-based superalloy V207 have been investigated.

## Experimental Methods
### 1. Samples

Three alloys were supplied by Rolls-Royce plc. for the present study, named V207G, V207J and V207K [15], and their nominal chemical compositions are given in Table 1. All alloys were prepared by argon gas atomisation followed by hot-isostatic pressing to produce cylindrical ingots. The cylindrical ingots were sectioned into discs using a Struers Accutom-50 (feed rate 0.005 and 3000 rpm), cleaned ultrasonically in an acetone/ethanol mixture and their surfaces ground with 1200 grade SiC paper. Surface preparation is known to affect oxidation behaviour, therefore 1200 grade paper was selected in order to make the surface finish comparable with previous work [16]. After grinding, the samples were ultrasonically cleaned in ethanol for a further 5 min before environmental exposures.

*Table 1: Nominal chemical composition in at% of the V207G, V207J and V207K alloys used in the present study* [15].

| At% | V207K | V207J | V207G |
|---|---|---|---|
| Ni | Bal. | Bal. | Bal. |
| Cr | 16 | 16 | 16 |
| Fe | 9 | 9 | 9 |
| Al | 5.5 | 5.5 | 5.5 |
| Co | 4 | 4 | 4 |
| Nb | 3.5 | 3.5 | 3.5 |
| Mo | 1.35 | 1.35 | 1.35 |
| Ti | 1 | 1 | 1 |
| **Mn** | **0** | **1** | **0** |
| **Si** | **0** | **0** | **0.5** |
| W | 0.9 | 0.9 | 0.9 |
| Ta | 0.7 | 0.7 | 0.7 |
| C | 0.15 | 0.15 | 0.15 |
| B | 0.15 | 0.15 | 0.15 |
| Zr | 0.035 | 0.035 | 0.035 |

### 2. Environmental exposures

A total of 12 samples were analysed through static furnace exposures during the current study. These comprised a separate sample of each of the 3 alloys exposed to the following 4 environmental conditions:

1. As-produced (initial microstructure)
2. Air
3. Air + 370 ppmv $SO_{2/3}$
4. Air + 370 ppmv $SO_{2/3}$ + sea salt.

The as-produced (control) samples of each alloy were heat treated in a Carbolite furnace, stabilised at 750°C, for 24 hours in quartz tubes evacuated to <$10^{-6}$ atm internal pressure then back-filled with Argon. Oxidised samples were subsequently produced by exposing each alloy to air in an open alumina crucible in a furnace stabilised at 750°C, for 24 hours. All samples were air-cooled after the furnace heat treatments.

Samples for sulfidation experiments were prepared by exposing the three alloys to air + 370 ppmv (by volume) $SO_2/SO_3$ mixture at 750°C in a furnace for 24 hours. The gas mixture was passed through a platinised honeycomb structure to catalyse the conversion from $SO_2$ to $SO_3$ and establish the equilibrium concentration of both. The importance of bringing laboratory-made gas mixtures to equilibrium through a heated catalyst bed has previously been shown by McAdam and Young [17], who found that Mn, heated to 800 °C and exposed to a gas mixture containing 22 vol.% $SO_2$ formed a combined scale of MnO and MnS when exposed to catalysed gas, whereas only MnO formed when exposed to non-catalysed gas. At the temperature of our tests, 750 °C, the reaction between $O_2$ + 370 ppmv $SO_2$ triggered by the catalyst is expected to form from an equilibrium (calculated) partial pressure of 176 ppmv $SO_3$.

Samples were prepared for hot-corrosion testing by pre-heating the surfaces to ~200 °C using a hot air gun, then spraying them with a saturated solution produced by dissolving sea salt in deionised water. The water evaporated on contact with the sample. The samples were weighed and imaged periodically using an optical microscope, to ensure the amount of salt coating the surface was 2.5 mg/cm$^2$ on each sample. The samples were then exposed to an air + 370 ppmv $SO_2/SO_3$ gas mixture at 750 °C for 24 hours (either in air or under the same conditions as the sulfidation experiment).

3. Microstructural Characterisation

X-ray diffraction (XRD) was used for phase identification. XRD was performed on all samples using a Bruker D8 Advance Gen.9 $\theta$–$2\theta$ diffractometer with Cu-k$\alpha$ radiation, operated at 40 kV and a current of 40 mA. Diffraction data were acquired between 20-100° $2\theta$, using steps of 0.02° and a dwell time of 7 s per step. The diffraction patterns were analysed by comparing them to patterns predicted using the Crystaldiffract software with crystallographic data obtained from the Inorganic Crystal Structure Database[1] (© FIZ Karlsruhe).

Samples for scanning electron microscopy (SEM) were cut using a Struers Accutom 50 saw and mounted in cold-setting conductive epoxy resin. They were then ground initially using successively finer grades of SiC paper, before polishing with diamond paste (3 $\mu$m-0.25 $\mu$m) and finally with a 0.04 $\mu$m colloidal silica suspension. Backscattered electron images were acquired, exploiting Z (atomic number) contrast in the unetched samples. A Zeiss GeminiSEM 300 SEM was used for imaging, operated between 5-15 kV with a beam current of 200 pA. To determine the compositions of the phases present, Energy Dispersive X-ray spectroscopy (EDX) was performed using an Oxford Instruments EDX detector and the data obtained was processed with Aztec software. The Cliff-Lorimer method was used to obtain semi-quantitative concentrations by utilising the Oxford Instruments built-in database to obtain k-factor values. Known peak overlaps (for instance Cr-L with O-K, Ta-M with Si-K) were deconvoluted either by referring to higher energy peaks (such as Ta-L) or by using the Oxford Instruments "TruMap" peak fitting and deconvolution algorithm. Oxide layer thicknesses were measured from backscattered electron micrographs using ImageJ software. Average values of the oxide layer thicknesses were obtained from 5 measurements per image, over 10 images. The associated errors were taken as the standard deviations over 20 equally spaced measurements from randomly sampled locations.

---

[1] icsd.cds.rsc.org

Results
1. Initial Microstructure

Phase identification in the as-produced material was initially performed through XRD. Diffractograms of the 3 alloys are shown in Figure 1. For all variations of the alloy the cubic Ni FCC-γ phase was identified, with a lattice parameter of 3.59 Å (obtained by matching all the available peaks between 2Θ=10-100˚). The alloying additions had no obvious effect on the lattice parameter of the γ phase. The presence of stable intermetallic Ni$_3$Nb δ phase and ordered superlattice L1$_2$ structure γ' Ni$_3$Al phase was also confirmed by XRD. The presence of manganese did not cause any additional phases to be detected by XRD. The presence of 0.5 wt% silicon, however, caused an increase in the intensity of the Bragg diffraction peaks associated with the δ phase. Reflections from common borides or carbides such as M$_{23}$C$_6$ (generally found at grain boundaries) or MC (bulk) were also not identified despite some of them being visible in the SEM micrograph in Figure 2, likely due to their low volume fraction being beyond XRD resolution.

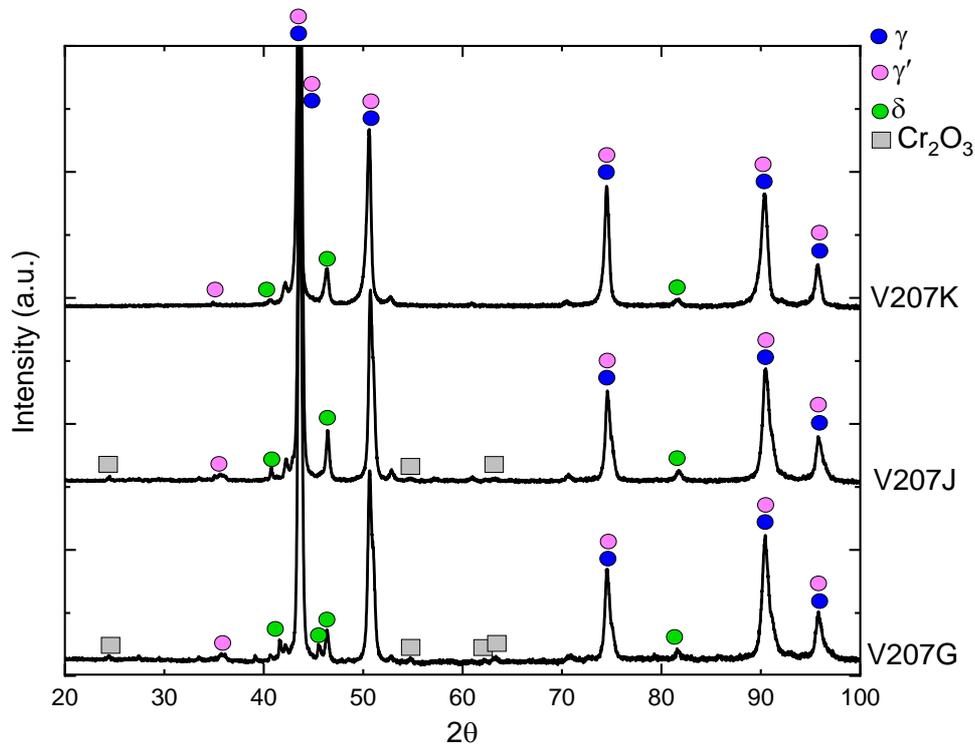

*Figure 1: X-ray diffractograms of the V207 K, J and G alloys in their as-hot isostatic pressed condition.*

Figure 2 shows backscattered electron imaging (BEI) micrographs of the V207 K, J and G alloys heat treated at 750°C for 24 hours under vacuum. Elongated features with light grey contrast, believed to be the δ intermetallic phase were visible in all the alloys. The V207G appears to have a larger fraction of smaller particles compared to the other 2 alloys, though detailed measurements are beyond the scope of this project. Carbides were also visible, spherical particles with a bright appearance in the BEI images, consistent with the appreciable refractory metal contents expected in these phases. The darker particles located at grain boundaries are postulated to be either carbides or borides, but their volume fraction was insufficient for XRD identification.

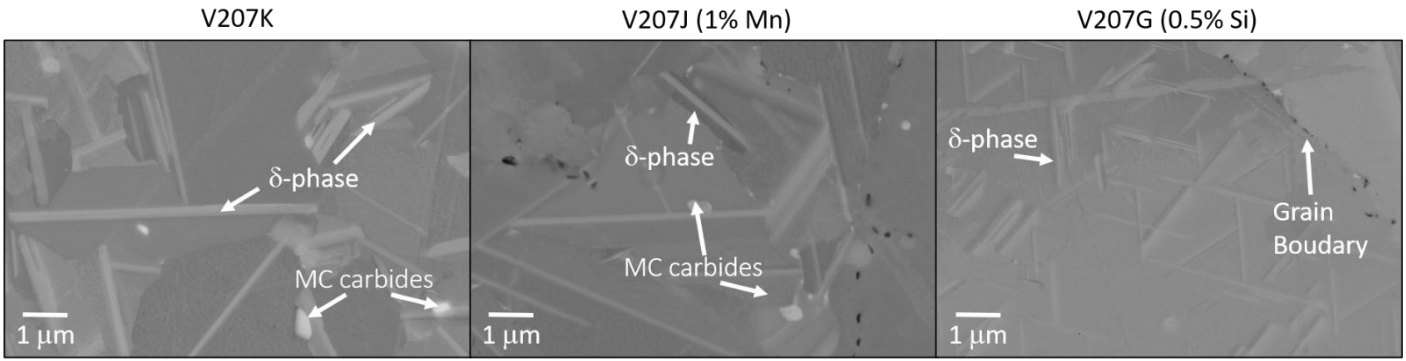

*Figure 2: BEI SEM micrographs of the V207 K, J, and G alloys after heat treatment at 750°C for 24 hours under vacuum.*

2. Oxidation

XRD was performed on the samples exposed in air for 100 hours for oxide phase identification (after 24 hours the oxide scale was too thin for accurate phase identification using a lab-based X-ray diffractometer). The X-ray diffractograms are shown in Figure 3. The presence of $Cr_2O_3$/$Al_2O_3$ was identified, but due to the two phases having the same (Corundum) crystal structure and similar lattice parameters, the corresponding peaks mostly overlapped and were therefore labelled as both. The presence of $MnCr_2O_4$ was confirmed on the V207J alloy, in addition to the $Cr_2O_3$/$Al_2O_3$ phases. This is consistent with previous work which showed through atom probe tomography that the surface oxide scale on this alloy after 100 hours at 800 °C was also a combined $MnCr_2O_4$/$Cr_2O_3$ scale [8]. After the 100-hour oxidation exposure the presence of γ-γ′ phases were also confirmed in the bulk alloys (as expected), though the heat treatment caused further precipitation of δ, shown by the increased relative peak intensity. No $NiO$, $TiO_2$, $SiO_2$, $Ta_2O_5$, $TiN$ or any oxides and spinels of Fe could be confirmed by XRD.

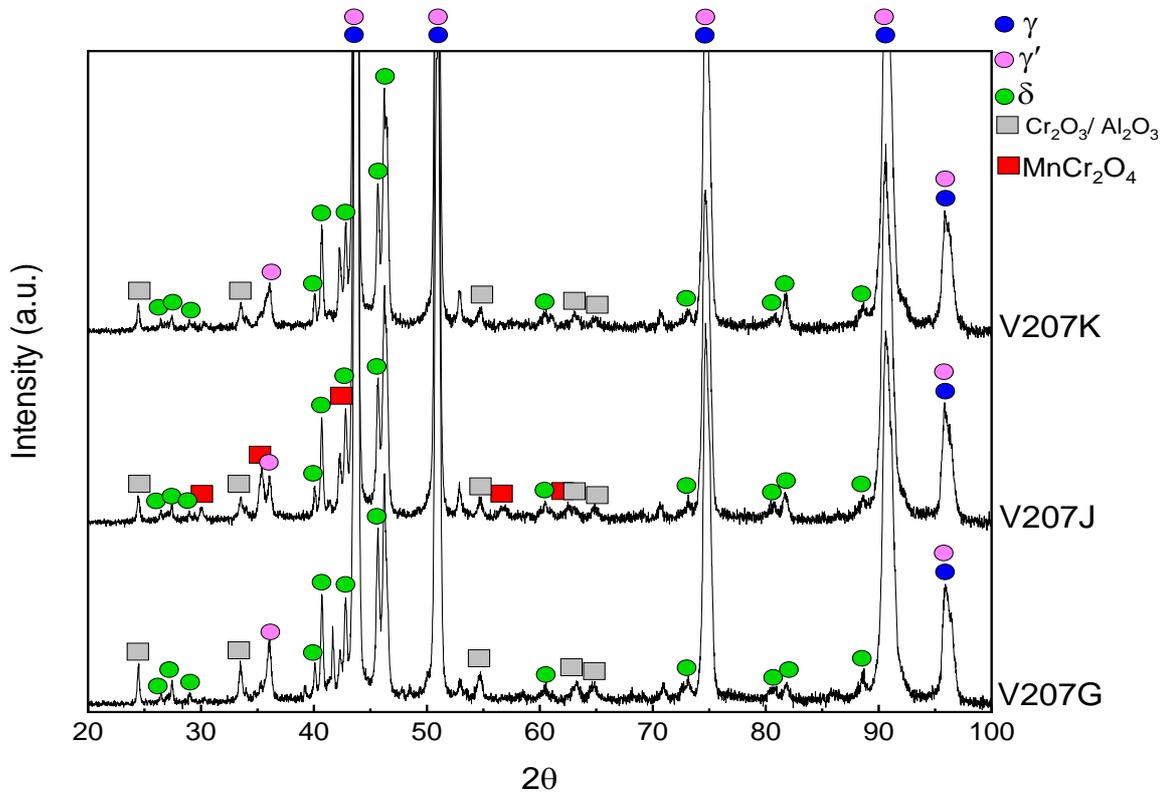

*Figure 3: X-ray diffractogram of the V207K, V207J (Mn-containing) and V207G (Si-containing) alloys after a 100-hour furnace heat treatment at 750˚C in air.*

Figure 4 shows backscattered SEM micrographs of the V207 K, J and G alloys, oxidised for (top) 24 hours and (bottom) 100 hours in air at 750°C. After 24 hours, the V207K and V207J alloys have an overall thinner oxide scale than V207G consisting of a continuous upper layer of $Cr_2O_3$ with inner $Al_2O_3$ intrusions. The addition of 1 wt% Mn did not appreciably alter the morphology or the thickness (within the error margin) of the oxide scale formed in 24 hours, though it is expected that the upper layer will be a combination of $Cr_2O_3$/$MnCr_2O_4$ in the V207J alloy and only of $Cr_2O_3$ in the V207K alloy, as previously established through atom probe tomography [18]. The presence of 0.5 wt% Si, however, significantly altered the morphology of the oxide scale formed on the V207G alloy. A continuous dual-layer chromia and alumina scale formed, with white nanoparticles (likely to be tantalum oxide, though beyond the resolution of the XRD in Figure 3) at the interface between the two layers. The overall damage depth is ~2x thicker than the V207K alloy after this short-term heat treatment, as shown in Table 2.

After 100 hours the morphology of each oxide scale on the V207K and V207J alloys remains unchanged but the thickness of the V207G (Si-containing) is 2-3x thinner than both other alloys due partly to a lack of sub-scale alumina intrusions, as shown in Figure 4 and Table 2. V207G alloy oxidises faster in the early stages, but the growth rate of the oxide scale is slower than the base V207K and Mn-containing V207J counterparts. It should be noted that though representative micrographs were selected, the oxide scales were inhomogeneous, as shown by the standard deviation values of the thickness of the oxide scales, shown in Table 2 (measured using ImageJ).

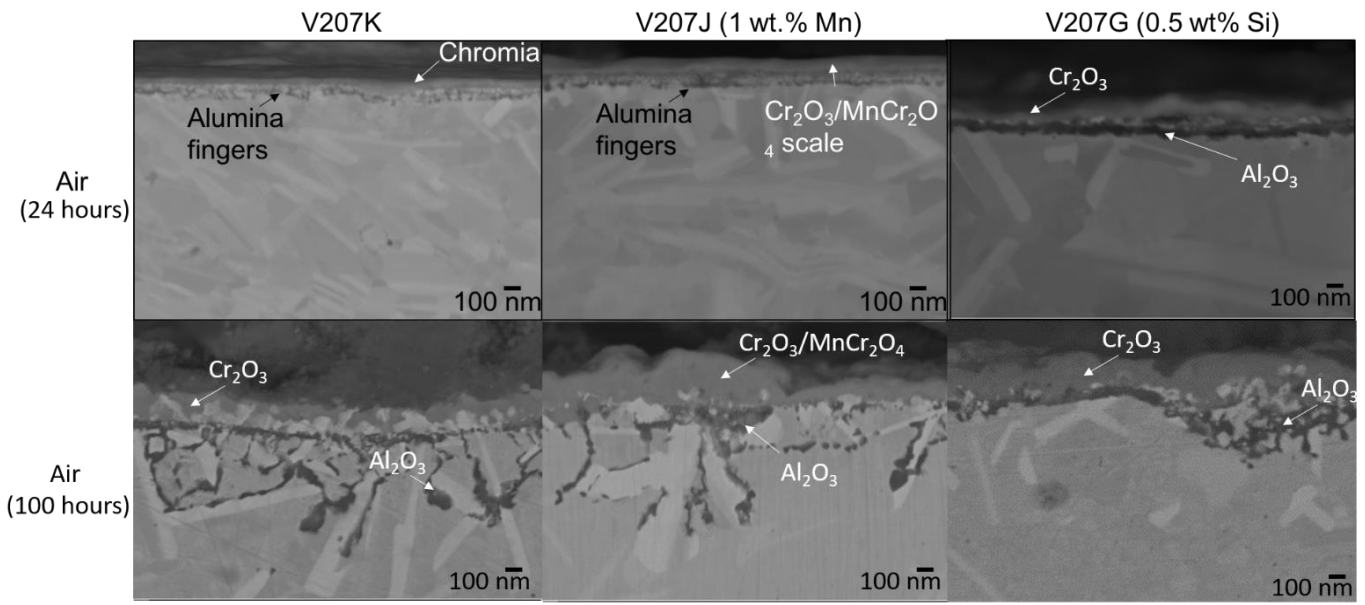

Figure 4: BEI SEM micrographs of the V207 K, J and G alloys, oxidised for (top) 24 hours and (bottom) 100 hours in air at 750°C.

Table 2: Oxide scale thicknesses measured from scanning electron micrographs (>20 measurements, randomly sampled, equally spaced, average ± standard deviation are shown).

| Sample | 24 hours Total oxide thickness (nm) | 100 hours Total oxide thickness (nm) |
|---|---|---|
| V207K | 110 ± 50 | 580 ± 380 |
| V207J (1 wt% Mn) | 120 ± 30 | 870 ± 350 |
| V207G (0.5 wt% Si) | 290 ± 160 | 370 ± 250 |

3. SO$_{2/3}$ exposure

X-ray diffractograms of the alloys exposed to air + 370 ppmv SO$_{2/3}$ showed that the combined oxide/sulfide scale after 24 hours was too thin to be identified within the resolution of the instrument, therefore these samples were characterised primarily though SEM, as shown in Figure 5.

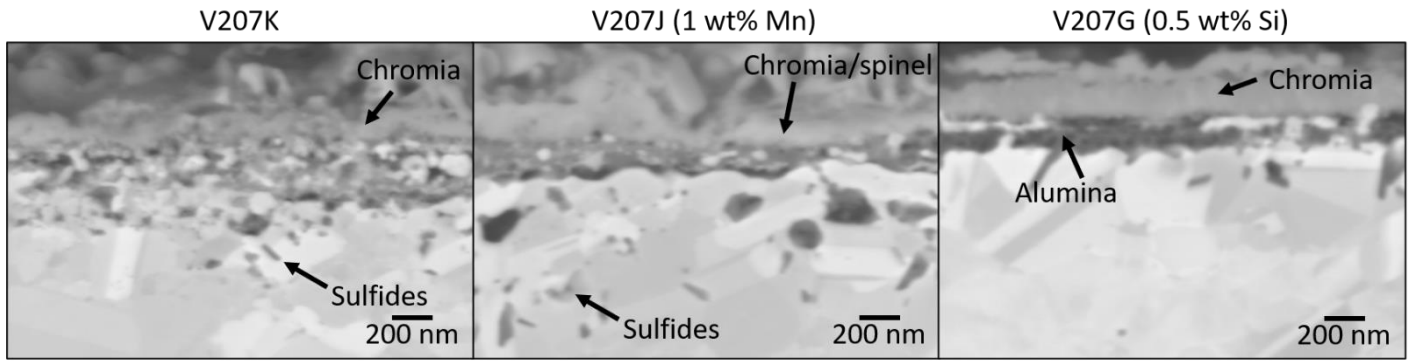

*Figure 5: BEI SEM micrographs of the V207 K, J and G alloys exposed to 370 ppmv SO$_{2/3}$ gas in air for 24 hours at 750°C.*

Figure 5 shows BEI micrographs of the samples exposed to air + 370 ppmv SO$_{2/3}$ for 24 hours at 750 °C. From the morphology and location of the surface damage, it was found that the V207K alloy has suffered more extensive damage from the oxygen/sulfur exposure than the equivalent sample oxidised in air for 24 hours, shown in Figure 4. The surface has a relatively thin, porous scale and internal alumina/sulfide formation beneath. The V207J alloy (Mn-containing) formed a thicker, less porous surface layer, though unlike for the V207K alloy, it retained enough integrity to reduce the extent of internal oxidation and alumina/spinel formation. It is noteworthy that due to the presence of Mn, the depth of damage has increased. The V207G alloy (Si-containing) exhibited the least surface damage under SO$_{2/3}$ exposure for 24 hours at 750°C, with a solid dual surface layer and no visible sub-scale intrusions. Some white particles are visible at the interface between the two surface layers in Figure 5, which are postulated to be tantalum oxide. The lighter appearance of these particles in backscattered imaging mode is indicative of heavy element presence, which according to the literature could be either tantalum oxide or a titanium-tantalum spinel [13], [19], [20]. Total thickness (damage depth) measurements are presented in Table 3 and show that the V207G alloys exhibited half the depth of damage compared to alloys V207J and K.

*Table 3: Surface damage thickness measured from scanning electron micrographs after 24 hours of environmental exposure to air + 370 ppmv SO$_{2/3}$ (>20 measurements, randomly sampled, equally spaced, average ± standard deviation are shown).*

| Sample | SO$_{2/3}$ exposure<br>Oxide + sulfide thickness (nm) |
|---|---|
| V207K | 800 ± 260 |
| V207J (1 wt% Mn) | 850 ± 290 |
| V207G (0.5 wt% Si) | 340±90 |

4. SO$_{2/3}$ and sea salt exposure

The phases formed through coupled SO$_{2/3}$ and sea salt exposure at 750 ˚C for 24 hours were confirmed by X-ray diffraction, shown in Figure 6. Diffraction peaks from the γ and γ′ phases were still present, though the thickness of the surface scale was sufficient to mask some of the bulk phases. Within the diffraction patterns, peaks from the NiO, FeCr$_2$O$_4$ spinel, TiO$_2$, and Cr$_2$O$_3$/Al$_2$O$_3$ phases were identified. The diffraction pattern obtained from alloy V207J also contained some evidence of MnS sulfides. The presence of silicon in the V207G alloy caused SiO$_2$ to be detected, although in extremely small amounts. The diffraction patterns from the V207J and V207G samples both contained peaks from Na$_2$SO$_4$, which was not part of the original alloy and was not added as a salt on the surface, therefore it is product of the reactions occurring during the heat treatment. A Bragg diffraction peak at 2θ=30.9° in the V207G alloy could not be identified from any of the expected phases.

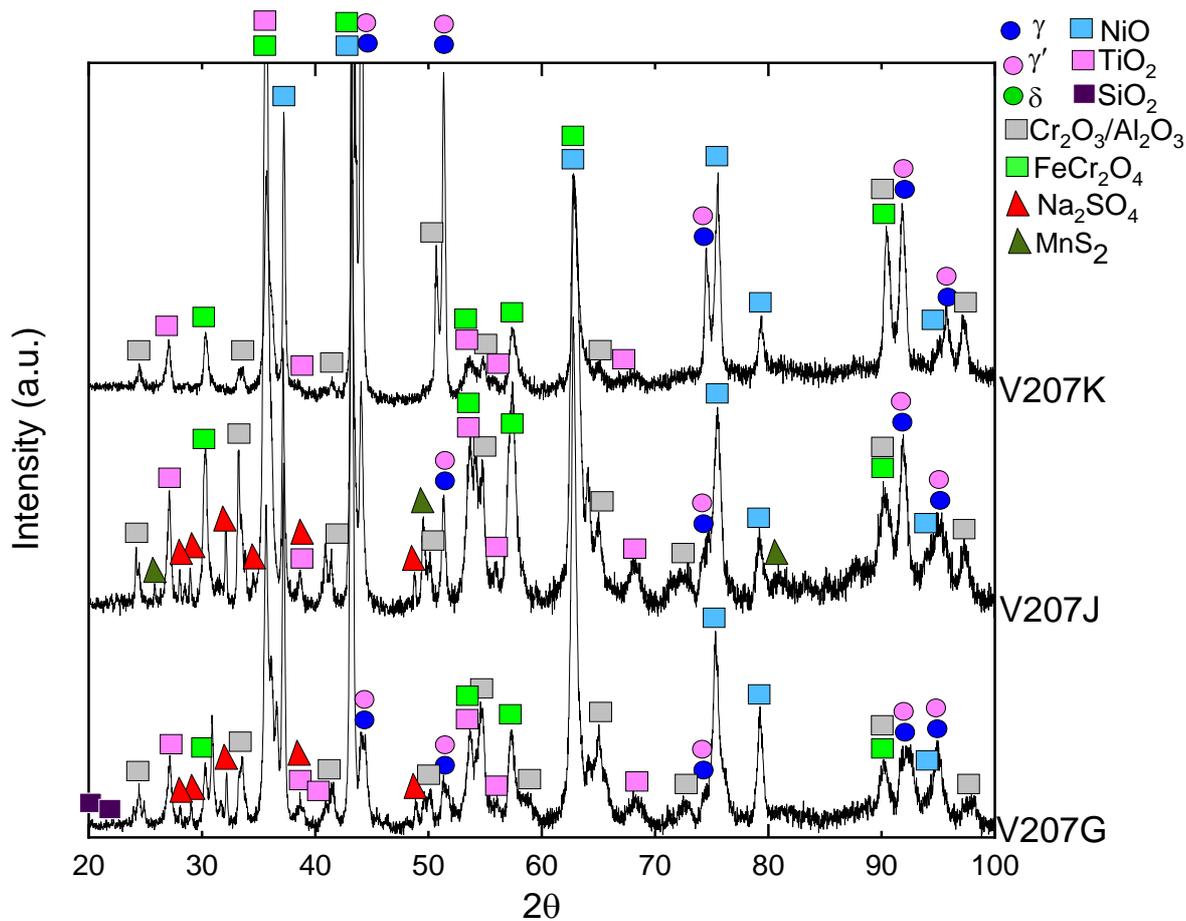

*Figure 6: X-ray diffractogram of the V207K, V207J (Mn-containing) and V207G (Si-containing) alloys after a 24-hour furnace heat treatment at 750˚C in air with 370 ppmv $SO_{2/3}$ and salt.*

Figure 7 shows BEI micrographs of cut and polished sections of the alloys after exposure to salt and 370 ppmv $SO_{2/3}$ in air at 750°C for 24 hours. The base alloy V207K was seen to have formed a thick uppermost oxide scale with severe spallation and visible evidence of oxide scale fluxing on the underlying substrate. Some internal oxidation and sulfidation was also observed beneath this external scale. The addition of 1wt% Mn did not visibly change the thickness or the spallation rate of the uppermost scale of the V207J alloy when compared to V207K. Despite this, the presence of Mn has visibly reduced oxide scale fluxing although it extended the overall damage depth (including the depth of internal oxidation/sulfidation). The V207G alloy (0.5 wt% Si) has the best resistance to environmental degradation under these conditions out of the 3 alloys analysed. However, the effect of the exposure on the surface scale is visible in the form of porosity (compared to Figures 4-5). Some internal oxidation and/or sulfidation underneath the surface scale was also visible, but to a lower extent compared to the V207K and V207J alloys. The surface scale is non-uniform, with wide variability captured by the damage depths mean and standard deviation values shown in table 6. More detailed SEM-EDX analysis of the phases formed on the surface of each alloy is shown in Figures 8-13.

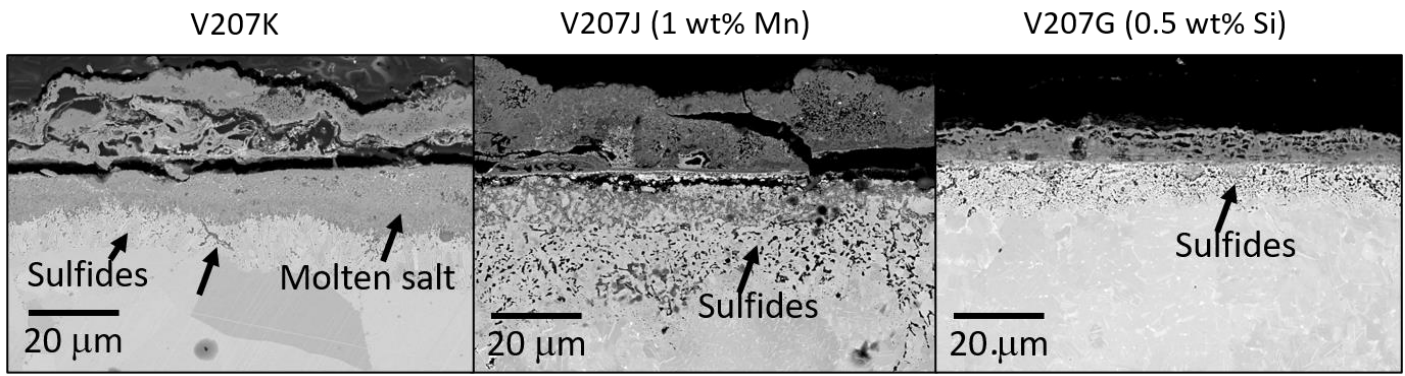

*Figure 7: BEI SEM micrographs of the V207 K, J and G alloys exposed to 370 ppmv $SO_{2/3}$ gas and sprayed with sea salt in air, for 24 hours, at 750°C.*

*Table 6: Surface damage thickness measured from scanning electron micrographs after 24 hours of environmental exposures (>20 measurements, randomly sampled, equally spaced, average ± standard deviation are shown). These measurements account for oxides/sulfides formation exclusively, and do not include any $\gamma'$ denuded zone which could be present, therefore extending the damage depth further.*

| Sample | Surface damage thickness ($\mu$m) |
|---|---|
| V207K | 22 ± 6 |
| V207J (1 wt% Mn) | 46.0 ± 8 |
| V207G (0.5 wt% Si) | 23 ± 7 |

A BEI SEM micrograph is shown in Figure 8 of the environmental damage on the baseline alloy V207K along with corresponding EDX elemental maps of the same region. The EDX data provided evidence of the location of the phases which were identified by XRD within the surface scale. In this case, the external scale is made of NiO, with an underlying $Cr_2O_3$ scale and the Fe-containing spinel $FeCr_2O_4$. There is evidence of dissolution of the chromia scale into the underlying alloy, extending the damage further when compared to the same alloy exposed without the salt, as shown in Figure 5.

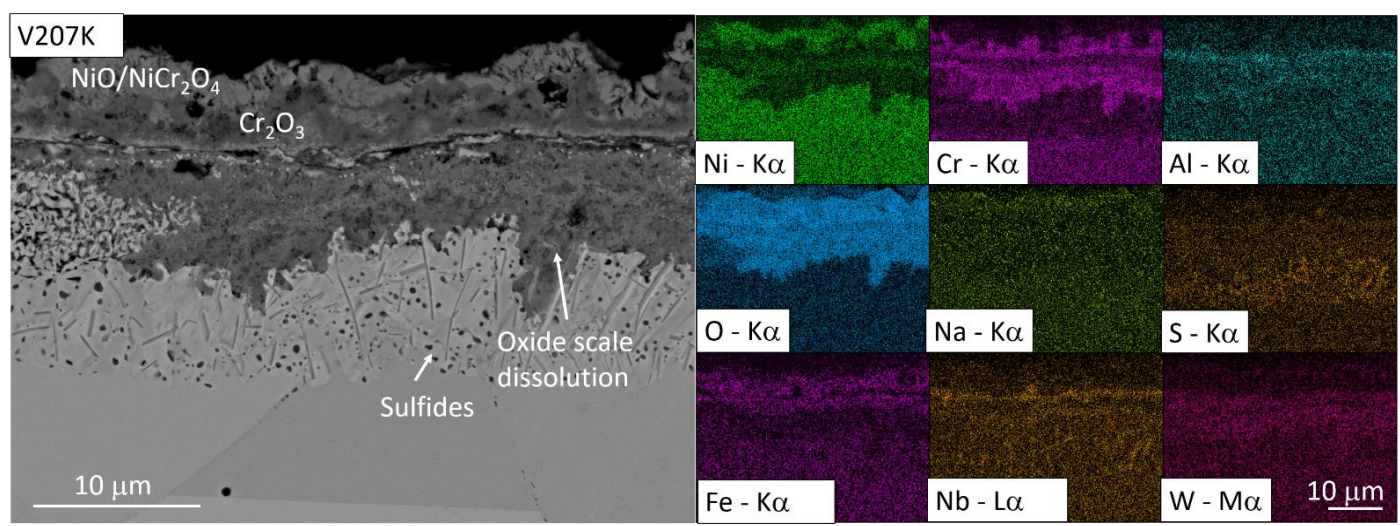

*Figure 8: BEI SEM micrograph and associated EDX elemental distribution maps of the V207K alloy sample exposed to air, 370 ppmv $SO_{2/3}$ and salt.*

Figure 9 shows a higher magnification image of the scale formed on the V207K alloy following exposure to air, with 370 ppmv $SO_{2/3}$ and salt, along with EDX elemental distribution maps of the region indicated by the red box. These data confirm the presence of a 10-20 μm denuded zone (Ni-rich and Cr-depleted in the EDX map) underneath the surface scale, which also contains evidence of internal oxidation and sulfidation. A distinct lack of alumina intrusions are observed compared to the oxidised sample (only one intrusion in the field of view). No evidence of chlorine was detected. Elongated, needle-like Nb sulfides have formed underneath the oxide scale. Niobium sulfides, much like most other refractory metal sulfides, have been previously reported to grow through sulfur ingress, which was speculated to be a result of them being anionic diffusers [21]. Their Cd(OH)$_2$-type crystal structure, made of two planes of sulfide ions with a plane of cations in between allows these structures to accommodate different metal-sulfur ratios, therefore allowing for a wide range of non-stoichiometric compositions to form [22], [23] depending on the local sulfur partial pressure [24], which can all be summarised using the general chemical formula $Nb_{1+\delta}S_2$. Nb has also been known to form carbo-sulfides [25]. The wide stoichiometric variation, coupled with the low volume fraction made the sulfides difficult to identify unambiguously in the X-ray diffraction patterns, Figure 6. Comparison with the SEM micrographs in Figure 8 and 9 shows the extent of non-uniformity of the surface scale, including localised spallation.

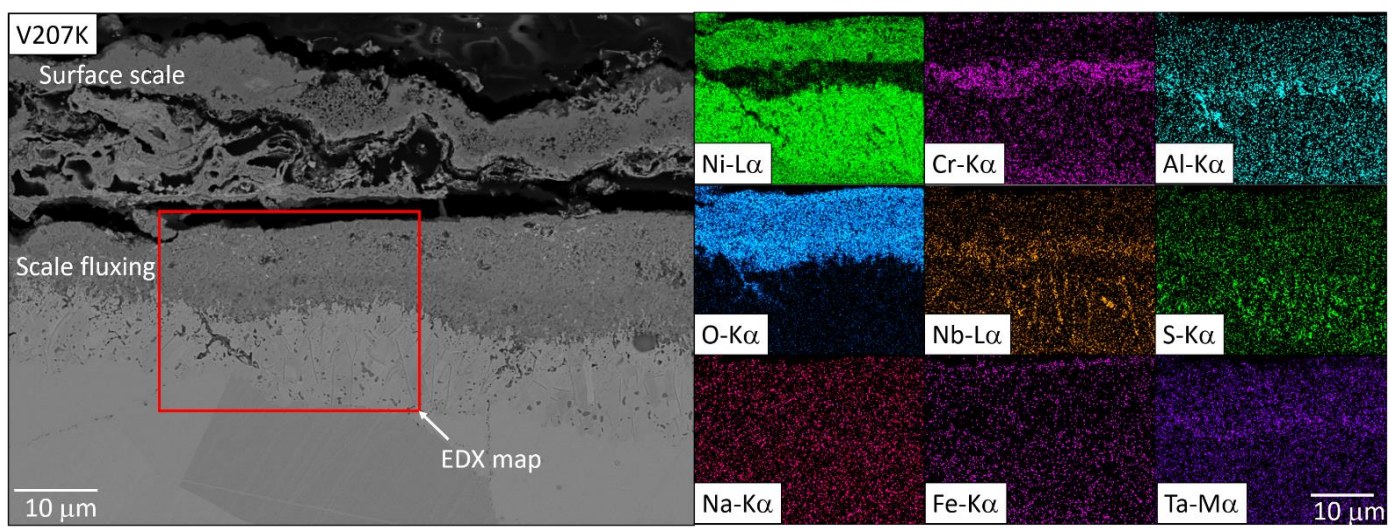

*Figure 9: BEI SEM micrograph and associated EDX elemental distribution maps showing a close-up detail of the V207K alloy sample exposed to air, 370 ppmv SO$_{2/3}$ and salt (overview show in in Figure 8).*

The SEM-EDX analysis of the V207J alloy (Mn containing) sample exposed to SO$_{2/3}$ and salt is shown in Figure 10. The MnCr$_2$O$_4$/Cr$_2$O$_3$ scale is visible, along with some Al$_2$O$_3$. In this alloy, Nb$_2$O$_5$ is visible in the surface scale, as well as an uppermost Fe-containing spinel, which the XRD data in figure 6 indicated should be FeCr$_2$O$_4$. Some evidence of oxide scale dissolution (fluxing) is visible in this sample as well, though not to the extent observed on the V207K alloy in Figures 8 and 9. Surface scale spallation is evident in Figure 10. The γ' denuded zone underneath the Cr$_2$O$_3$/MnCr$_2$O$_4$ scale is enriched in Ni. Manganese sulfides formed underneath the Al$_2$O$_3$ intrusions, shown more in detail in Figure 11. The damage depth was almost double that of the V207K alloy, due partly to the presence of manganese sulfides. In this alloy, no evidence of niobium sulfides was observed. Na$_2$SO$_4$ was unambiguously identified in this sample using the X-ray diffractogram in Figure 6, which according to the EDX map in Figure 10 is likely to be located towards the outer surface of the oxide scale.

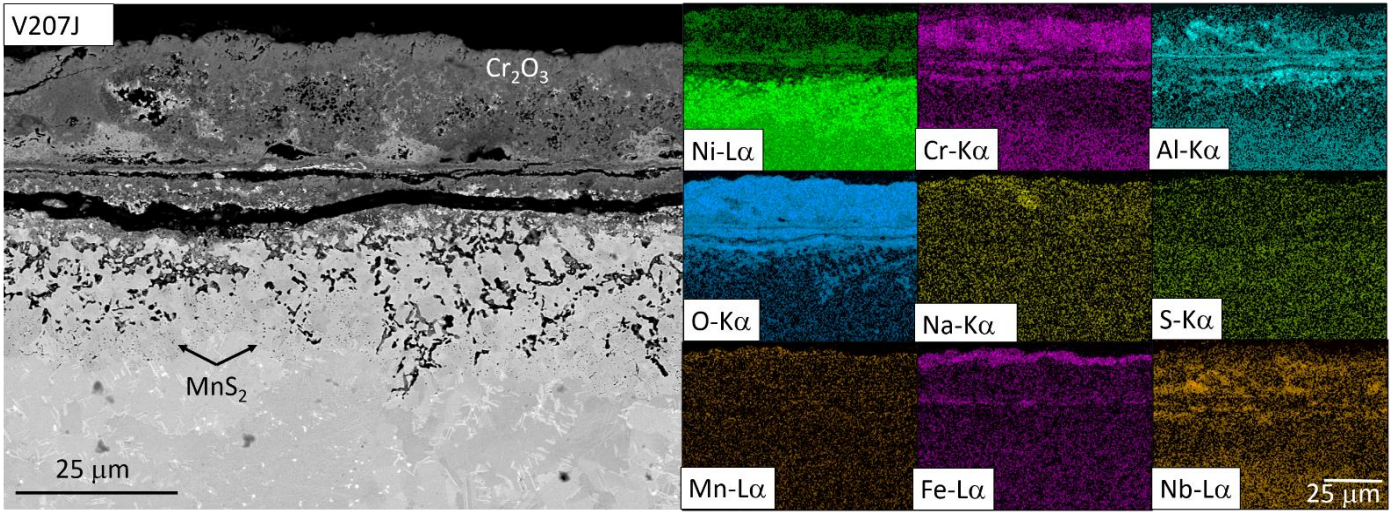

*Figure 10: BEI SEM micrograph and associated EDX elemental distribution maps of the V207J alloy sample exposed to air, 370 ppmv $SO_{2/3}$ and salt.*

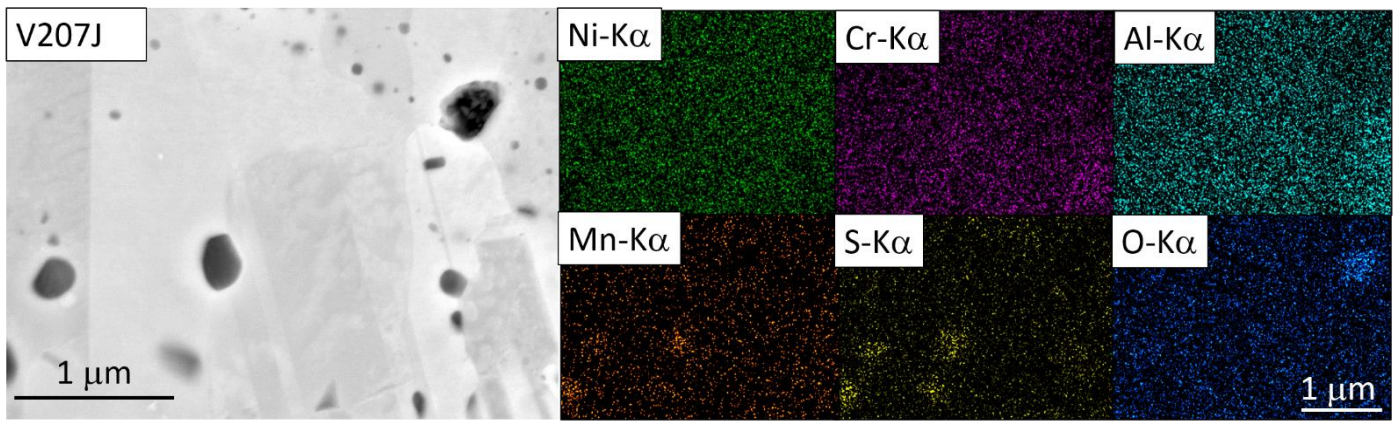

*Figure 11: BEI SEM micrograph and associated EDX elemental distribution maps showing a close-up detail of the V207J alloy sample exposed to air, 370 ppmv $SO_{2/3}$ and salt (overview show in in Figure 10).*

Figure 12 shows a backscattered SEM micrograph and associated EDX maps of the V207G (Si-containing) alloy exposed to $SO_{2/3}$ and salt for 24 hours at 750°C. The surface scale consisted of a dual-layer chromia-alumina scale, though the chromia was interspersed with $FeCr_2O_4$ spinel. Nb was observed in the form of distinct particles at the oxide-metal interface, which appear Si-rich (Figure 13), possibly Nb silicides, though their presence could not be confirmed by XRD. Figure 14 shows a higher magnification view of the area underneath the surface scale, within the γ' denuded zone. The Si and O elemental maps indicated that $SiO_2$ formed as intrusions underneath the oxide scale. Niobium sulfides were also observed, though only lower amounts of smaller niobium sulfide particles than in the V207K alloy, with a Ni-enriched zone (γ' denuded zone) surrounding them.

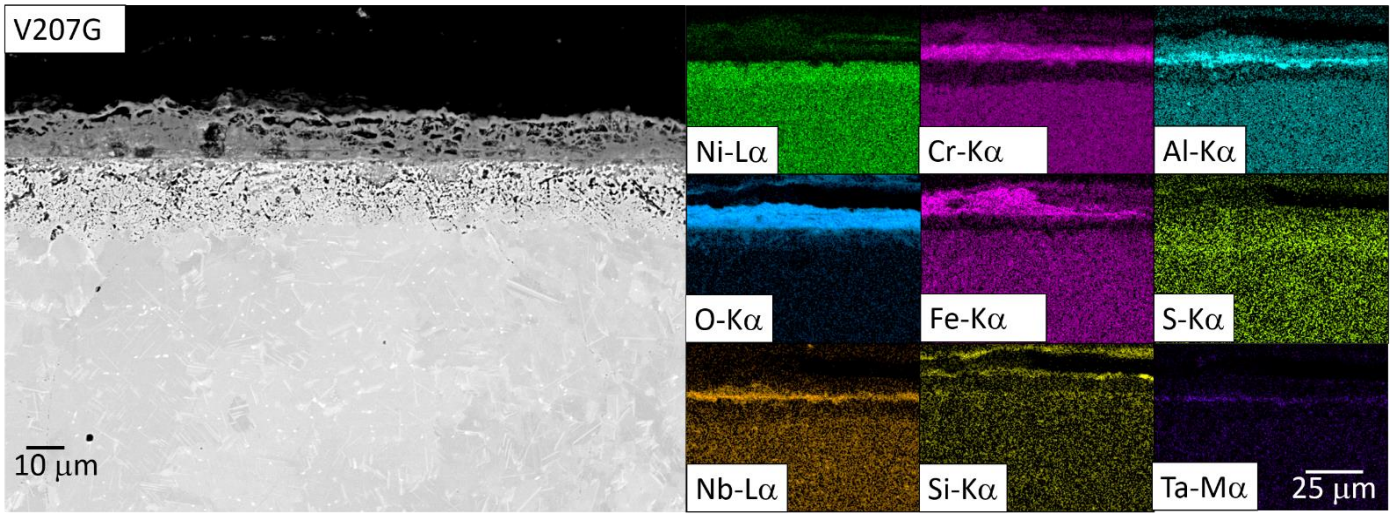

*Figure 12: BEI SEM micrograph and associated EDX elemental distribution maps of the V207G alloy sample exposed to air, 370 ppmv $SO_{2/3}$ and salt.*

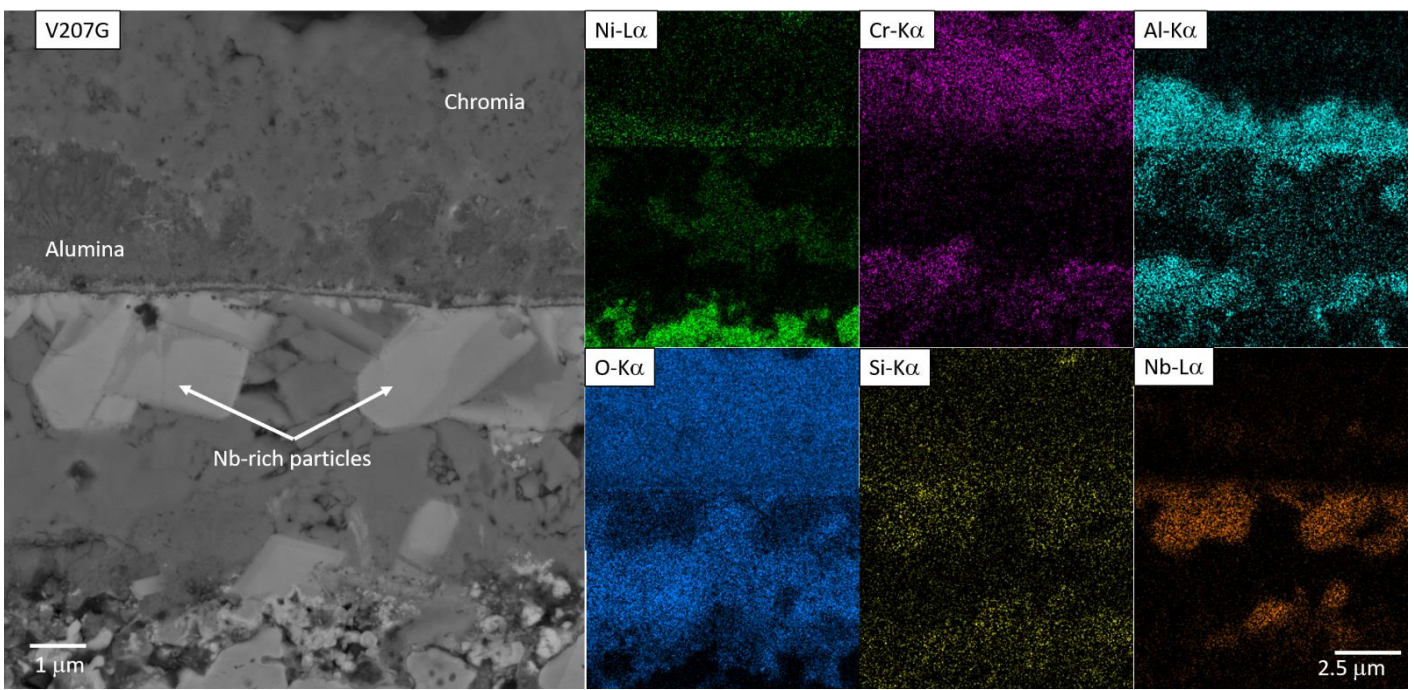

*Figure 13: BEI SEM micrograph and associated EDX elemental distribution maps of the interface between the inward growing and outwards growing oxides, showing the presence of Nb-rich particles.*

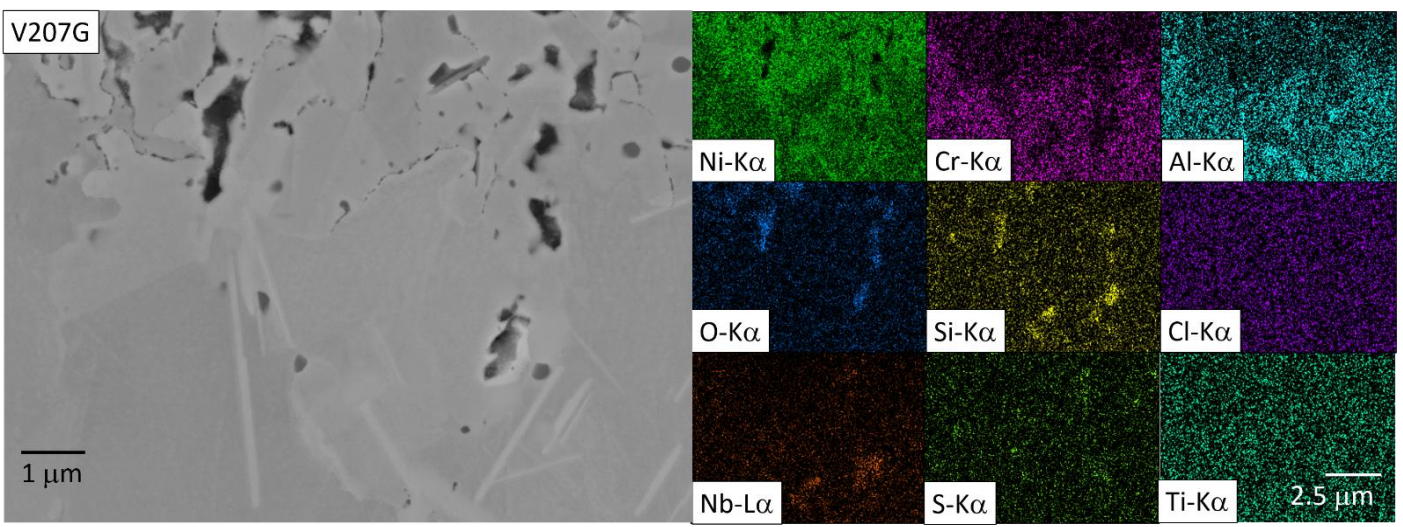

*Figure 14: BEI SEM and associated EDX elemental distribution maps showing a close-up detail of the internal oxidation and sulfidation products found on the V207G alloy sample exposed to air, 370 ppmv SO$_{2/3}$ and salt (overview show in in Figure 12).*

5. Environmental effect summary:

A summary of the phases formed from each environmental exposure is shown in Table 5.

*Table 5: overview of phases present in each sample after environmental exposures, observed by XRD and SEM-EDX.*

|  | | Environment: | | | | | | | | |
|---|---|---|---|---|---|---|---|---|---|---|
|  | | Inert gas HT (24h @750°C) | | | Air (100h @750°C) | | | Air +300ppm SO$_{2/3}$+sea salt (24h @750°C) | | |
|  | **Phase** | **V207K** | **V207J** | **V207G** | **V207K** | **V207J** | **V207G** | **V207K** | **V207J** | **V207G** |
| Bulk | $\gamma$-Ni | ✓ | ✓ | ✓ | ✓ | ✓ | ✓ | ✓ | ✓ | ✓ |
| | $\gamma'$-Ni$_3$Al | ✓ | ✓ | ✓ | ✓ | ✓ | ✓ | ✓ | ✓ | ✓ |
| | $\delta$-Ni$_3$Nb | ✓ | ✓ | ✓ | ✓ | ✓ | ✓ | - | - | - |
| Oxides | Cr$_2$O$_3$/Al$_2$O$_3$ | - | ✓ | ✓ | ✓ | ✓ | ✓ | ✓ | ✓ | ✓ |
| | MnCr$_2$O$_4$ | - | - | - | - | ✓ | - | - | - | - |
| | NiO | - | - | - | - | - | - | ✓ | ✓ | ✓ |
| | FeCr$_2$O$_4$ | - | - | - | - | - | - | ✓ | ✓ | ✓ |
| | TiO$_2$ | - | - | - | - | - | - | ✓ | ✓ | ✓ |
| | SiO$_2$ | - | - | - | - | - | - | - | - | ✓ |
| | Ta$_2$O$_5$ | - | - | - | - | - | - | - | - | - |
| Sulfides/ Sulfates | MnS$_2$ | - | - | - | - | - | - | - | ✓ | - |
| | NbS$_2$ | - | - | - | - | - | - | - | - | - |
| | Na$_2$SO$_4$ | - | - | - | - | - | - | ✓ | ✓ | ✓ |

Discussion: thermodynamic, kinetic and electrochemical reaction mechanisms

Part 1: Oxidation

During furnace oxidation testing, the metallic sample surfaces were exposed to air under atmospheric pressure at 750 °C. With a constant oxygen supply to the surface, the partial pressure of oxygen beneath the surface is expected to follow an error function diffusion profile according to Fick's law (rather than a parabolic profile), with different oxide phases forming once the required oxygen partial pressure ($p_{O_2}$) is reached. Diagram 15 shows an Ellingham diagram containing the oxygen partial pressure and Gibbs free energy of formation of each of the oxides which were observed on the V207 series alloys and how their stability varies with temperature. Diagram 15 was adapted from a variety of databases and publications, which were used to calculate the partial pressure using the equation $\Delta G = RTln(p_{O_2})$ where G is the Gibbs free energy, R is the gas constant, P is the partial pressure and T is the absolute temperature. Each reaction was normalised by number of moles of oxygen, in order to get comparable values for each type of oxide.

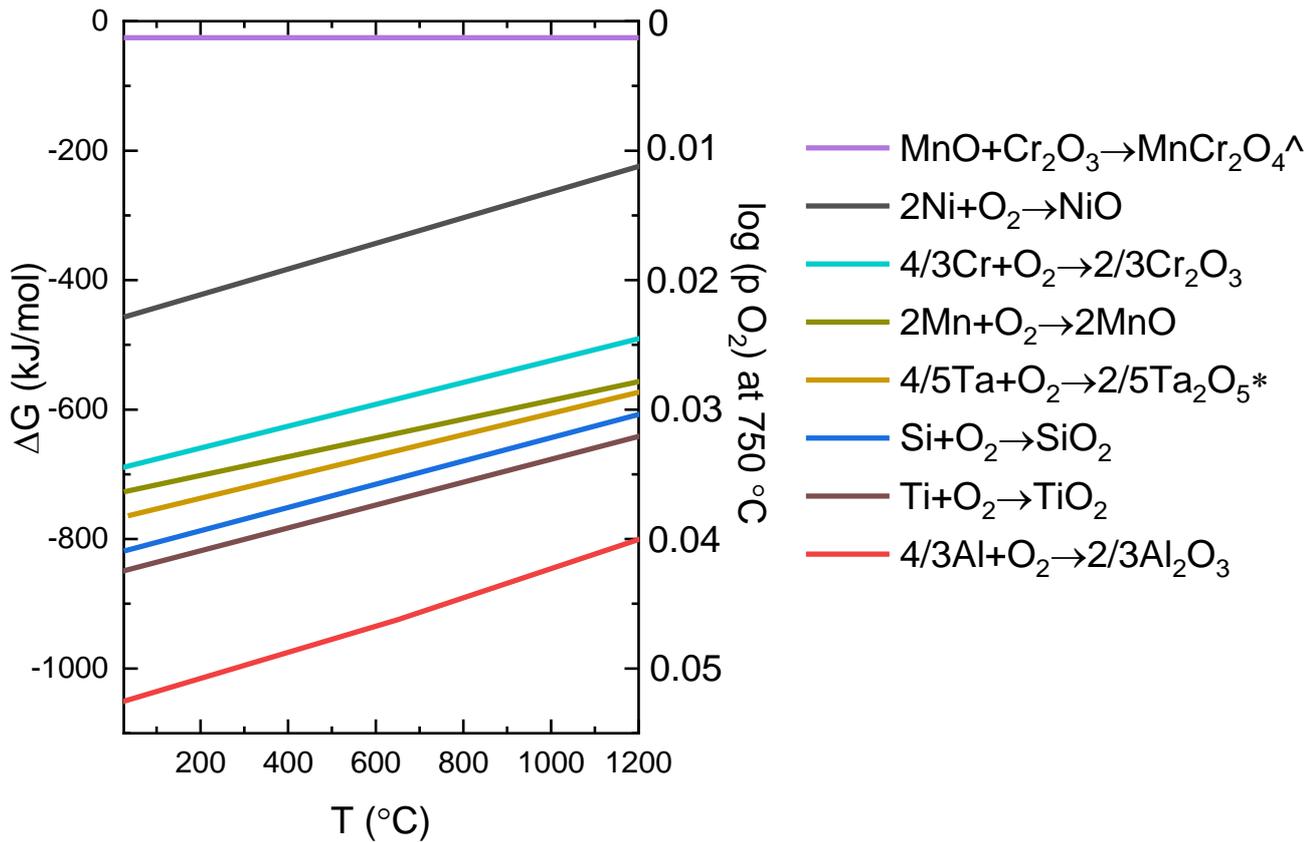

*Diagram 15: Ellingham diagram of oxide formation, taken and adapted from a variety of databases [26], [27] and journal papers [28].*

Based on the alloy baseline composition (V207K), according to diagram 15 a multi-layered scale of NiO, $Fe_2O_3$, $Cr_2O_3$, $Al_2O_3$ may be expected to form on the V207K alloy. However, due to kinetic considerations, not all of the thermodynamically stable oxides and spinels were observed and only chromia and alumina were identified. NiO has a tendency to form in the earliest stages of oxidation, due to the relative abundance of nickel in the alloy compared to other alloying elements allowing Ni to have a shorter mean free path of diffusion towards the surface [29]. However, NiO is a relatively less stable oxide, therefore the growth of chromia and alumina quickly supersedes its formation as they may form at lower partial pressures of oxygen [30]. The V207J (Mn-containing) alloy also formed $MnCr_2O_4$ spinel, in addition to $Cr_2O_3$ and $Al_2O_3$ (as observed in a previous study [31]), whereas the V207G (Si containing) alloy formed some $SiO_2$ which could be detected by XRD but were not located in the SEM-EDX analysis. Low concentrations of Si added to other nickel-based superalloys (0.5 wt% or less) have previously been shown to cause the localised formation of silica nanoparticles as opposed to a continuous scale [13]. The nanoparticles were located between the $Cr_2O_3$ and $Al_2O_3$, a location which is consistent with the partial pressures of formation predicted by the Ellingham diagram in diagram 15. In some alloys, silicon has also been shown to promote the formation of a continuous layer of alumina, as opposed to disconnected alumina intrusions found underneath the surface scale [13], [32], [33].

Part 2: $SO_{2/3}$ exposure

The chromia scale that forms on the surface of the V207K, J and G alloys, while protective in air, is permeable to $SO_{2/3}$. Previous studies on pre-oxidised chromia forming alloys have shown that the partial pressure of $SO_{2/3}$ present influences the response of the chromia scale: at partial pressures >$3\times10^{-7}$ atm sulfur will fail to penetrate the material, by reacting with the outer surface to form a scale of $Cr_3S_4$ [34]. Others, such as De Asmundis *et al*. [35] found that exposure of chromia forming alloys to $SO_{2/3}$ at 700-1000°C led to the formation of a chromia scale with no visible sulfidation. The primary issue with these earlier studies is a lack of information regarding whether the gas mixture was brought to the equilibrium partial pressure of $SO_{2/3}$ using a catalyst or not, therefore making the studies non-comparable. The importance of the addition of a catalyst was shown in more recent studies [17].

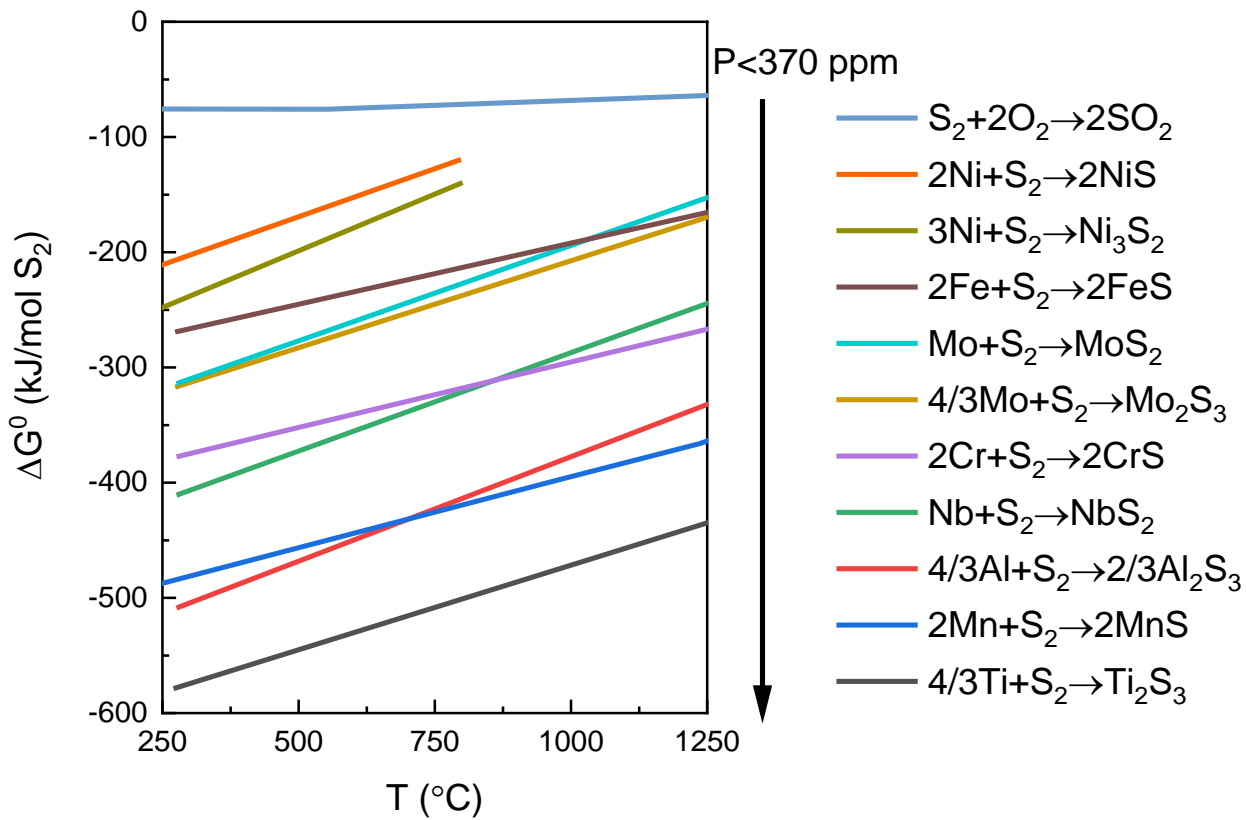

**Diagram 16:** *Ellingham Diagram for selected sulfides, normalised by mol $S_2$. Data sources: $SO_2$ [36], NiS [37], $Ni_3S_2$ [37], FeS [38], $MoS_2$ [38], $Mo_2S_3$ [38], CrS [39], $NbS_2$ [36], $Al_2S_3$ [38], MnS [37], $Ti_2S_3$ [38] .*

In the case of the V207J alloy, the added Mn caused the formation of $MnCr_2O_4$ on the surface as well as $Cr_2O_3$, resulting in a mixed oxide scale. Previous studies reported that Mn in the presence of $SO_{2/3}$ will form MnO on the surface and MnS intrusions underneath [17]. According to the Ellingham diagram in Diagram 16, MnS requires the lowest partial pressure of $SO_{2/3}$ to form, therefore they form at a greater depth in the alloy. The Mn content of the V207J alloy allows for deeper damage due to the formation of MnS, but lower damage extent, due to the remaining surface protection afforded by the presence of MnO.

In the case of V207G alloy, the silicon content caused the formation of a continuous dual layer of $Cr_2O_3$ and $Al_2O_3$. The combination appears passivating under the current exposure conditions and the superalloy surface was not exposed to further attack.

### Part 3: $SO_{2/3}$ and sea salt exposure
***Surface scale:***
After exposure to $SO_{2/3}$ and sea salt, characterisation of the surface scale confirmed the presence of NiO, $FeCr_2O_4$, $Na_2SO_4$ with low amounts of $Cr_2O_3$/$Al_2O_3$ and $TiO_2$.

The addition of sea salt (compared to the sulfidation experiment) promoted the following reaction steps: the NaCl also reacts with the $SO_3$ and $H_2O$, to create hydrochloric acid and sodium sulfate (shown by XRD in Figure 6) according to Reaction A below.

**Reaction A:** $2NaCl_{(s)} + SO_{3\,(g)} + H_2O_{(l)} \rightarrow Na_2SO_{4\,(s)} + 2HCl_{(aq)}$

Reaction A is well documented in the scientific literature since the 1940s as the starting point of NaCl corrosion in the presence of $SO_3$, as described both in Young's book "High Temperature Oxidation and Corrosion of Metals" (chapter 8, equation 8.11) [38] and Kofstad's book "High Temperature Corrosion" (chapter 13, equation 13.2) [40]. Previous studies, such as those by Rapp [5] elucidated thermodynamic mechanisms for the dissolution of the oxide scales by $Na_2SO_4$. Fused $Na_2SO_4$ was treated as an ionic liquid that separated into $Na_2O$ and $SO_3$ according to the following reaction: $Na_2SO_{4(s)} \leftrightarrow SO_{3(g)} + Na_2O_{(s)}$, which would then lead to the dissolution of oxides though an acid-base reaction with $SO_3$ acting as the acid and $Na_2O$ as the base [5]. $Na_2SO_4$ has a melting point of 884 °C, though it can form a low-melting point eutectic with $T_m$=687 °C in the presence of $NiSO_4$ [41]. The absence of this low melting point mixed sulfate salt in the current experiments can be explained using thermodynamic arguments. The Nernst equation for the dissociation reaction is:

$$\Delta G^0 = RTln(\frac{a_{Na_2SO_4}}{a_{Na_2O}\, p_{SO_3}})$$

where (G is the Gibbs free energy, R is the gas constant, T is absolute temperature, a is the activity of the chemical species listed and p is the partial pressure). Similarly, the Nernst equations of several potential reactions was plotted in Diagram 17 as a phase stability diagram for Na-S-O compounds, also overlayed with the phase stability diagram for Ni-S-O compounds. Collectively these diagrams serve to explain the phases observed in the current study. The experimental parameters of the current study are in within the window of simultaneous thermodynamic stability of both $Na_2SO_4$ and NiO, indicated in red in Diagram 17. It is therefore unsurprising that both were observed as part of the surface scale. $Na_2O$ is only stable at much lower partial pressures of $SO_3$ than the experimental parameters used and $NiSO_4$ forms at substantially higher pressures of $SO_3$, therefore none of the low-melting point mixed salts were expected to form in the current experimental set-up.

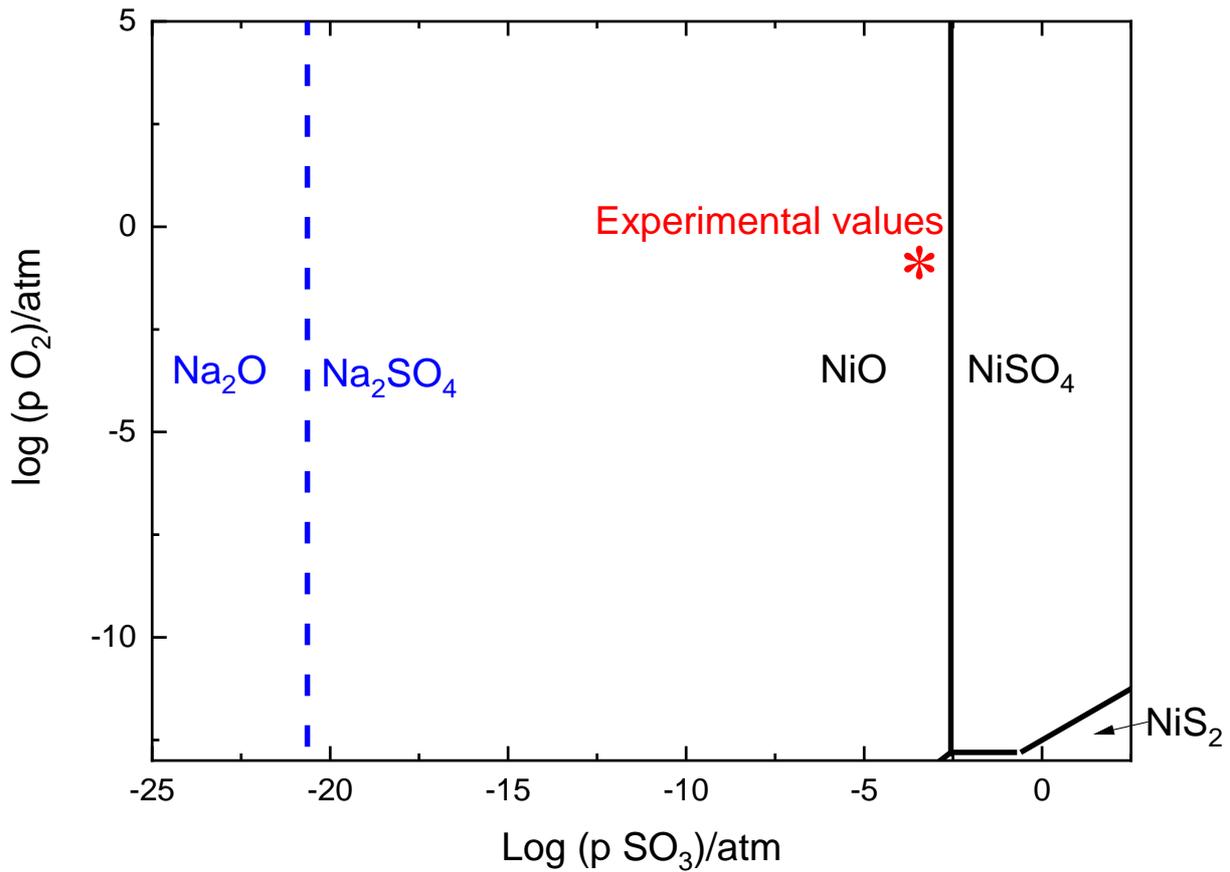

*Diagram 17: Phase stability diagram of Ni-S-O (black solid lines) overlayed with the phase stability diagram of Na-S-O compounds (blue dashed lines) at 750 °C and in red an indication of the experimental parameters used in the present study. Data obtained and adapted from [40].*

Chromia was also detected as part of the surface scale after the salt exposures, however, the relative amount and its morphology were indicative of a partial dissolution. Chromia is an amphoteric oxide, which dissolves both through acidic and basic fluxing mechanisms. Reaction A is the likely cause of acidic fluxing and dissolution of the chromia scale, leading to reactions B and C below:

**Reaction B:** $Cr_2O_{3(s)} + SO_{3(g)} \rightarrow Cr^{3+}{}_{(aq)} + SO_4{}^{2-}$ [40]

Reactions B leads to the $Cr_2O_3$ scale dissolution but does not explain the absence of the $Al_2O_3$ layer, when compared to oxidized (non-corroded) samples. Previously, Hwang and Rapp [42] and Gheno *et al* suggested synergistic dissolution mechanism for chromia and alumina forming nickel-based superalloys, which link the dissolution of both oxides as cause and effect of each other [43]. The equation in reaction C is based on their acidic dissolution work, however, in this case the presence of hydrochloric acid adds a further reaction forming aluminium chloride:

**Reaction C:** $Al_2O_{3(s)} + 6HCl_{(g)} \rightarrow 2AlCl_{3\ (g)} + 3H_2O_{(g)}$

Reaction C can occur at temperatures as low as 400°C. Critically, $AlCl_3$ is gaseous at the temperature of the present study, 750°C (and at any temperature above ~180°C), explaining the fact that chlorine could not be located in any of the samples using EDX in Figures 8-12. This reaction did not, however, cause the total dissolution of the alumina, but it could if the amount of NaCl applied to the surface was increased.

*Internal sulfidation:*

The rate at which an alloy fails through surface scaling far exceeds the rate of internal sulfidation, therefore this phenomenon is seldom studied. In the current experiments, most of the sulfur formed $Na_2SO_4$, therefore the partial pressure of sulfur that remains available for internal sulfidation is reduced. Niobium sulfides ($Nb_{1+\delta}S_2$) were observed in the V207G and K alloys, in the form of elongated particles beneath the surface oxide scale, whereas in the V207J alloy, manganese sulfides (MnS) were observed instead. No other internal sulfides were identified after coupled $SO_{2/3}$ and sea salt exposure.

The reactions explained in Reactions A-C allowed the dissolution of the primary protective surface oxide scales, leaving the surface open to potential further chemical attack. The $SO_{2/3}$ then diffused through the scale into the alloy to form sulfides beneath the scale. Diagram 16 shows an Ellingham diagram correlating potential sulfides which could form through reactions with the alloying elements of the V207 alloys and their relative thermodynamic stabilities. An arrow has been included to indicate the conditions of the current experiments (partial pressure was calculated using $\Delta G = RT ln(p_{S_2})$ at T=750 °C). From a purely thermodynamic perspective, Diagram 16 would predict the formation of Ni, Fe and Mo after experimental exposures, as well as $Cr_2S_3/Al_2S_3$, $Nb_{1+\delta}S_2$ and (in the V207J alloy) MnS. The formation of $Na_2SO_4$ is enough to cause a reduction in partial pressure of $SO_{2/3}$, therefore reducing the amount available for the formation of sulfides.

In accounting for sufide formation, kinetic effects and growth mechanisms also need to be taken into account. Both Ni and Fe-based sulfides tend to grow outwards as part of the surface scale. In previous work, Nb has been shown to be highly beneficial to the sulfidation rate of some alloys, by reducing the rate of sulfidation of Fe [44] and Ni [45], which both normally form through fast outwards growth rather than as internal oxides [40], [46]. The suppression of $Ni_xS_y$ formation is particularly beneficial since they form low melting point eutectics, such as the Ni-S sulfide indicated in Diagram 17, which is liquid above 635°C [47].

Another similar experiment was performed on a Ni-Cr-Al based turbine disc alloy (RR1000) performed at 750°C in air +300 ppmv $SO_{2/3}$ and $NaCl-Na_2SO_4$ salts (both under static furnace exposures and under dynamic fatigue testing conditions) and in the absence of Nb, Ti-rich Cr sulfides formed instead [48], [49]. It is possible that Nb suppresses the formation of $Cr_3S_2$ as well, though more studies will be required to prove this. Nb has also been shown to reduce the sulfidation rate of Al [44] both in binary and ternary alloys. MnS is extremely stable and slower growing than any other transition metal sulfide, due to its unusual cubic structure ($\alpha$-MnS) which does not allow for large deviation from stoichiometry [50]. Although it has a similar growth rate to chromia, its presence is not generally considered beneficial for the corrosion resistance of alloys because it does not form a stable surface scale.

In the V207G alloy no $SiO_2$ was observed after oxidation testing or exposure to $SO_{2/3}$ alone, presumably due to its presence in the form of nanoparticles at the $Cr_2O_3/Al_2O_3$ interface (as expected from the Ellingham diagram in diagram 15), making them harder to identify due to their size and location. After salt exposure, the dissolution of both $Cr_2O_3$ and $Al_2O_3$ as outlined in reactions A-C above, allowed direct observation of silica nanoparticles, as seen in Figure 12. According to previous studies, silica, though unlikely to be present in sufficient amounts to form a continuous surface scale, is the only one of the oxides formed on this alloy which would not be dissolved through acidic (or basic) dissolution mechanisms, an observation which explains their presence in the V207G alloy after sulfur and salt exposure [5].

## Conclusions

The present study assessed the oxidation, sulfidation and hot corrosion response at 750 °C of three variants of the polycrystalline superalloy V207: a baseline alloy, one containing 1 wt% Mn and one containing 0.5 wt% Si. The following conclusions were drawn:

- After the 24-hour heat treatment at 750 °C in air, the baseline alloy formed an oxide scale comprising chromia, with alumina intrusions visible underneath. Longer exposures caused a thickening of the oxide scale and alumina intrusions, but no alterations to the morphology.
- The addition of 0.5 wt.% Si did not alter the oxidation rate when compared to the baseline sample after 24 hours, however, it changed the oxide morphology. Silicon additions promoted the creation of a thin, continuous alumina layer beneath the surface chromia scale. The different scale morphology reduced the oxidation rate in long-term experiments (100 hours).
- The addition of 1 wt.% Mn improved oxidation resistance by creating $MnCr_2O_4$ instead of the chromia both after 24 and 100 hours, which reduced the oxidation rate but did not change the oxide scale morphology and alumina intrusion morphology.
- The more compact dual-layer chromia-alumina scale produced through the addition of silicon reduced the sulfidation damage depth by 2/3, after samples were exposed to air + 370 ppmv $SO_{2/3}$ for 24 hours. The presence of Mn (a known sulfur scavenger) proved less effective, reducing the damage extent but increasing the overall depth.
- The silicon-containing alloy proved more resistant to hot corrosion from NaCl + 370 ppmv $SO_{2/3}$ combined exposure than the Mn-containing alloy or the baseline alloy, reducing the damage depth by ½ after 24 hours at 750 °C.
- A thermodynamic argument based on phase stability was presented to explain the oxidation, sulfidation and hot corrosion mechanisms.

## Acknowledgements

The Rolls-Royce EPSRC strategic partnership grant EP/M005607/1 is acknowledged for part-funding this study, along with EPSRC fellowship EP/S013881/1, and in-kind contributions from the Royal Academy of Engineering in the form of an associate research fellowship. Rolls-Royce plc provided samples and some funding.

## Data Availability

Data contained in this manuscript will be made available by the corresponding author upon reasonable request.